\documentclass[showpacs,preprint]{revtex4}

\usepackage{graphicx}
\usepackage{dcolumn}
\usepackage{bm}
\usepackage{amsfonts,amsmath,amssymb}
\usepackage[squaren]{SIunits}
\makeatletter
\begin{document}

\title{Magnetic and Transport Properties in Gd$_{1-x}$Sr$_x$CoO$_{3-\delta}$ ($x$= 0.10-0.70)}
\author{X. G. Luo,  H. Li, X. H. Chen\footnote{Corresponding author. \emph{Electronic address:}
chenxh@ustc.edu.cn}, Y. M. Xiong, G. Wu, G. Y. Wang, C. H. Wang,
W. J. Miao, and X. Li} \affiliation{Hefei National Laboratory for
Physical Science at Microscale and Department of Physics,
University of Science and Technology of China, Hefei, Anhui
230026, People's Republic of China\\}
\date{Received          }

\begin{abstract}
Magnetic and transport properties of polycrystalline
Gd$_{1-x}$Sr$_x$CoO$_{3-\delta}$ ($x$= 0.10-0.70) annealed under
the oxygen pressure of 165 atm  at 500 $\celsius$  are
systematically investigated . Cluster-glass behavior is observed
in the low doping range, while a ferromagnetic transition occurs
at the high doping level. Transport measurements indicate
insulator-like behavior for the samples with $x \leq$ 0.30, an
insulator-metal (IM) transition around $x$ = 0.35, and metallic
behavior for higher $x$ samples.  However, the large oxygen
deficiency leads to a reentrance of insulator-like state for the
samples with $x \geq$ 0.60. Annealing procedure under the high
oxygen pressure at high temperature can diminish the oxygen
deficiency and leads to restore the metallic state. The small
radius of Gd$^{3+}$ ion results in the less conduction, and lower
$T_{\rm c}$ compared to La$_{1-x}$Sr$_x$CoO$_{3}$ and
Nd$_{1-x}$Sr$_x$CoO$_{3}$ due to the large structural distortion
and the stability of low-spin state.
\end{abstract}


\maketitle

\section{INTRODUCTION}

To understand the peculiar electromagnetic properties of the
perovskite-type cobalt oxides, Ln$_{1-x}$A$_x$CoO$_3$ (Ln=rare
earth element, A=alkli earth metal), such as the large negative
MR,\cite{Brinceno,Golovanov,Mahendiram} spin-(or cluster-)glass
magnetism,\cite{Itoh,Goodenough,Mukherjee,Kumar} spin-state
transition\cite{Mukherjee1,Lashkareva} and insulator-metal
transition induced by doping or temperature,\cite{Shingo,Moritomo}
numerous works have been performed by many researchers. One
important part of these works is to change Ln$^{3+}$ or A$^{2+}$,
in order to get information of the electronic structure and
magnetic states with different ionic radii or hole
concentrations.\cite{Itoh,Goodenough,Saitoh,Sehlin,Ganguly,Yoshii}

One striking feature for the richness of the physical properties
of Ln$_{1-x}$A$_x$CoO$_3$, compared to other transition oxides
like CMR manganites, nickelates and cuprates, is the presence of
the various spin states for trivalent cobalt ions (low-spin LS,
Co$^{\rm \@Roman{3}}$: $t_{2g}^6e_{g}^0$; intermediate spin IS,
Co$^{\rm \@roman{3}}$: $t_{2g}^5e_{g}^1$; high spin HS, Co$^{3+}$:
$t_{2g}^4e_{g}^2$) and tetravalent cobalt ions (LS, Co$^{\rm
\@Roman{4}}$: $t_{2g}^5e_{g}^0$; IS, Co$^{\rm \@roman{4}}$:
$t_{2g}^4e_{g}^1$; HS, Co$^{4+}$: $t_{2g}^3e_{g}^2$), and the
relative narrow energy gap between these spin states. This makes a
thermally spin-state transition occur
easily.\cite{Goodenough,Raccah} Recent experimental and
theoretical investigations indicate that the spin states are LS
and the mixture of IS/LS for tetravalent and trivalent cobalt
ions,
respectively.\cite{Lashkareva,Saitoh,Korotin,Yamagushi,Louca,Kobayashi,Zobel,Ravindran}
The conversion of different spin-states arises from the
competition between comparable in magnitude the crystal-field with
energy $\Delta_{\rm CF}$ ($t_{\rm 2g}$-$e_{\rm g}$ splitting) and
the intraatomic (Hund) exchange with energy $J_{\rm ex}$, leading
to the redistribution of electrons between $t_{\rm 2g}$ and
$e_{\rm g}$ levels. $\Delta_{\rm CF}$ is found to be very
sensitive to the variation in the Co-O bond length ($d_{\rm
Co-O}$), so the subtle balance between $\Delta_{\rm CF}$ and
$J_{\rm ex}$ may be easily disrupted by different kinds of effect,
such as the hole-doping and the chemical/external
pressure.\cite{Asai,Lengsdorf,Vogt,Fita} Among them, chemical
pressure on CoO$_6$ octahedra is usually generated by decreasing
the average Ln-site ionic radius $<r>$, which can cause the
insulating nonmagnetic LS state because of the increase of the
$\Delta_{\rm CF}$ with the reduction of the CoO$_6$ octahedra
volume, which results in the depopulation of the magnetic $e_{\rm
g}$ level.

Another pronounced feature in perovskite cobaltites is that the
ferromagnetic (FM) state evolves as a result of increasing hole
doping level in low $x$, in the paramagnetic matrix with dominant
antiferromagnetic (AFM) superexchange interactions between
Co$^{3+}$ ions through a spin- or cluster-glass-state
region.\cite{Wu,Nam} The competition between FM and AFM
interactions leads to a highly inhomogeneous ground state
exhibiting the coexistence of FM regions, spin-glass regions and
hole-poor LS regions.\cite{Lashkareva,Kunhs,Hoch,Ghoshray} The
evolution of these regions and the spin-states with hole
concentration leads to the intricate magnetic and electronic
behaviors. Studies on La$_{1-x}$Sr$_{x}$CoO$_3$ system (La(Sr)
compounds) reveal a rather rich magnetic and electronic phase
diagram with the doping level: spin-glass for $x <$ 0.18, cluster
ferromagnetic behaviors for $x \geq$ 0.18, insulator-like/metallic
resistivity, metal-insulator transition at $x \approx$ 0.20, and
so on.\cite{Itoh,Wu} Ca- and Ba-doped compounds
La$_{1-x}$A$_{x}$CoO$_{3}$ (A= Ca, and Ba) have also been studied
intensively.\cite{Rao,Kriener}

Besides the La compounds, other important perovskite cobaltites
like Ln$_{1-x}$Sr$_x$CoO$_3$ (Ln$^{3+}$ = Pr$^{3+}$, Nd$^{3+}$,
Sm$^{3+}$, Eu$^{3+}$, Gd$^{3+}$ etc.) also exhibit complex
magnetic and electrical
properties.\cite{Rao,Brinks,Fandado,Senaris1,Stauffer,Ryu,Krimer}
In this paper, we investigated Sr-doped gadolinium cobaltites
systematically. However, most of the work on
Gd$_{1-x}$Sr$_{x}$CoO$_{3-\delta}$ (Gd(Sr) compounds) was focused
on the evolution of crystal structure with Sr
doping,\cite{Ryu,Takeda} magnetic and transport properties in
relative high temperature ($T
>$ 77 K).\cite{Ryu,Takeda} The Gd$^{3+}$ ion has different
characters from La$^{3+}$ ion, for example, smaller ionic radius
than La$^{3+}$ ion, the high magnetic moment with no (L-S)
anisotropy (L = 0, S = 7/2),\cite{Cabezudo} in contrast to the
nonmagnetic La$^{3+}$ (L = 0, S = 0). Therefore, the contrasting
behavior to La(Sr) compounds should be expected in Gd(Sr)
compounds. In fact, some distinct properties from La(Sr) compounds
have been reported in the Gd(Sr) compounds. In undoped sample
GdCoO$_3$, the cobalt ions are in low-spin state below 270
K,\cite{Casalot} while in LaCoO$_3$, the mixture of LS/IS state is
observed above about 100 K.\cite{Sudheendra,Chang,Kzinek}
Especially, no metallic behavior is reported in
Gd$_{1-x}$Sr$_x$CoO$_{3-\delta}$ system below 300 K so
far,\cite{Ryu,Takeda,Cabezudo} in contrast to the metallic
resistivity for $x >$ 0.20 in La(Sr) compounds. In this paper,
samples annealed under the oxygen pressure of 165 atm at 500
$\celsius$ are systematically investigated. Cluster-glass behavior
is found in low doping range, while a ferromagnetic transition is
observed at high doping level. Transport measurements indicate
insulator-like behavior for the samples with $x \leq$ 0.30, an
insulator-metal (IM) transition around $x$ = 0.35, and metallic
behavior for the samples with higher $x$. The system reenters
insulator-like state for $x \geq$ 0.60, resulting from the
existence of large oxygen deficiency, and metallic state can be
restored by annealing the samples at 900 $\celsius$ and under the
high oxygen pressure of 240 atm. We found that the Gd(Sr)
compounds possess less conductance and lower FM transition
temperature relative to La(Sr) compounds\cite{Wu} and
Nd$_{1-x}$Sr$_{x}$CoO$_3$ (Nd(Sr) compounds),\cite{Stauffer} which
is attributed to the large structural distortion and the stability
of low-spin state arising from the smaller radius of Gd$^{3+}$
ion.

\section{EXPERIMENT}

Polycrystalline Gd$_{1-x}$Sr$_x$CoO$_{3-\delta}$ ($x$= 0.10-0.70)
samples were prepared through conventional solid-state reaction.
The stoichiometric amounts of Gd$_2$O$_3$, SrCO$_3$, and
Co$_3$O$_4$ powders were thoroughly mixed and fired at 1200
$\degree$C. After that, the mixture was reground and pressed into
pellets which were sintered at 1200 $\celsius$ for 24 h. This
procedure was repeated for three times. In order to get the
homogeneous sample with less oxygen deficiency, the samples  were
then annealed under the oxygen pressure of 165 atm at 500
$\celsius$ for 48 h. X-ray diffraction (XRD) was performed by
Rigaku D/max-A X-Ray diffractometer (XRD) with graphite
monochromated CuK$\alpha$ radiation ($\lambda$ = 1.5418 ${\rm
\AA}$) at room temperature. Magnetization measurement was carried
out with a superconducting quantum interference device (SQUID)
magnetometer (MPMS-7XL, Quantum Design). The resistivity
measurements were performed by using the standard ac four-probe
method. The magnetic field was supplied by a superconducting
magnet system (Oxford Instruments). We also determined the oxygen
content of the samples using K$_2$Cr$_2$O$_7$ titration method. An
appropriate amount of sample (about 30 mg) was dissolved in the
mixture of vitriol and phosphate acid, then the high valent Co
ions were deoxidized to divalent ones with Fe$^{2+}$ ions, and
finally the excess Fe$^{2+}$ ions were titrated with
K$_2$Cr$_2$O$_7$ solution.

\section{EXPERIMENTAL RESULTS}

\subsection{XRD patterns}

Figure 1 shows the XRD patterns for
Gd$_{1-x}$Sr$_x$CoO$_{3-\delta}$ ($x$= 0.10-0.70). The XRD
patterns indicate that the obtained samples were all single phase
and exhibited an O-type orthorhombic GdFeO$_3$-like distorted
perovskite structure (SG: Pnma, $a_{\rm ort}/\sqrt{2} \leq c_{\rm
ort}/2 \leq b_{\rm ort}\sqrt{2}$), which has been used in the
GdCoO$_3$ \cite{Casalot} and
Gd$_{0.5}$Sr$_{0.5}$CoO$_3$.\cite{Vanitha} The variation of
lattice parameters with $x$ was plotted in Fig. 2. The lattice
parameters $a$ and $c$ increase monotonically with the doping
level, while $b$ firstly decreases with $x<0.4$, then increases
with further increasing $x$. It is found that the difference among
the reduced parameters becomes less with increasing $x$, and
nearly the same for $x=0.7$, indicative of a tendency of cubic
symmetry. It is reasonable since SrCoO$_3$ is
cubic.\cite{Shaplygin} The lattice volume, shown in the inset of
Fig. 2, increases monotonously with $x$, which is consistent with
the substitution of the larger Sr$^{2+}$ ions ($^{\rm
\@Roman{12}}r_{\rm Sr^{2+}}$= 1.58 ${\rm \AA}$) for the smaller
Gd$^{3+}$ ions ($^{\rm \@Roman{12}}r_{\rm Gd^{3+}}$= 1.22 ${\rm
\AA}$ ).\cite{Shannon,Shannon1}

\subsection{Magnetic properties}

Figure 3 shows the molar magnetic susceptibility $\chi_{\rm m}(T)$
as a function of temperature in zero-field cooled (ZFC) and
field-cooled (FC) procedures for the four typical samples. For the
sample with $x$ = 0.60, the $\chi_{\rm m}(T)$ curve behaves as a
typical ferromagnet that a pronounced increase in $\chi_{\rm
m}(T)$ occurs below a temperature $T_{\rm c}$ ($\approx 180 K$),
with a paramagnetic signal at low temperature, which comes from
the contribution of magnetic Gd$^{3+}$ ions. The $T_{\rm c}$
corresponds to the ferromagnetic transition temperature. With
decreasing Sr concentration, the rise becomes to slow down and
$T_{\rm c}$ decreases sharply (from about 120 K for $x$ = 0.45 to
80 K for $x$ = 0.10). A rounded maximum appears below $T_{\rm c}$
in the ZFC $\chi_{\rm m}(T)$ curves. This behavior may be ascribed
to the spin-glass (SG) magnetism, which has been reported in
La(Sr) and Nd(Sr) compounds with low doping
level.\cite{Goodenough,Wu,Stauffer} In La(Sr) and Nd(Sr)
compounds, the ZFC curve shows a cusp at the freezing temperature
of SG.\cite{Goodenough,Wu,Stauffer} However, the ZFC and FC curves
in the two compounds bifurcate at a temperature much higher than
that of the cusp,\cite{Goodenough,Wu,Stauffer} in contrast to the
behavior observed in present Gd(Sr) compounds. Therefore, one may
assume that Gd(Sr) compounds have different magnetic property from
La(Sr) and Nd(Sr) system in the low doping region. In order to
clarify it, the frequency dependence of ac susceptibility was
measured. Figure 4(a) and 4(b) shows the temperature dependence of
$\chi_{\rm m}^\prime(T)$ and $\chi_{\rm m}^{\prime\prime}(T)$ (the
in-phase and out-of-phase component of the ZFC ac susceptibility)
for $x$ = the sample with 0.10 taken at 10 and 1000 Hz. A peak can
be observed in both $\chi_{\rm m}^\prime(T)$ and $\chi_{\rm
m}^{\prime\prime}(T)$ at 75 K. In La(Sr) and Nd(Sr) compounds, the
temperature corresponding to the peak $\chi_{\rm m}^\prime(T)$ is
the same as that of cusp in ZFC dc
susceptibility.\cite{Wu,Stauffer} However, this temperature is
much higher by 30 K than that corresponding to the maximum of the
the ZFC $\chi_{\rm m}(T)$ in Gd$_{0.9}$Sr$_{0.1}$CoO$_{3-\delta}$.
It is well known that the peak temperature in ac susceptibility
for a spin-glass should be the same as that in ZFC susceptibility,
which represents a magnetic freezing temperature of SG. Thus the
ac susceptibility data exclude the possibility of the existence of
a spin glass in Gd(Sr) compounds with low doping. The absence of
the spin glass for $x$ =0.10 is illustrated more clearly by the
nearly frequency independence of the peak temperature of
$\chi_{\rm m}^\prime(T)$ in Fig. 4(c), which shows a "closeup" of
this peak in $\chi_{\rm m}^\prime(T)$, measured with a temperature
spacing of 0.5 K. Measured from 1 Hz to 1000 Hz, the peak
temperature of $\chi_{\rm m}^\prime(T)$ shifts to higher side by
less than 0.65$\%$ (=0.5K/77K), while it is much larger than this
value in La(Sr) and Nd(Sr) compounds (for instance, it changes
about 2$\%$ in La(Sr) compounds).\cite{Wu,Stauffer} Consequently,
the peak in ac susceptibility could not be an indication of a SG
behavior. Fig. 4(d) and 4(e) also displays the "closeup" of the ac
susceptibility $\chi_{\rm m}^\prime(T)$ for the samples with $x$ =
0.30 and 0.60. With the frequency changing from 1 Hz to 1000 Hz,
the peak of $\chi_{\rm m}^\prime(T)$ for $x$ = 0.30 shifts to
higher temperature by less than 0.5 K, while the peak position for
$x$ = 0.60 is independent of the frequency. From Fig. 3(a) and
3(b), it can be concluded that there is a small ferromagnetic
component in the samples with $x$ =0.10 and 0.30. Apparently,
ferromagnetic clusters are very diluted by nonmagnetic matrix in
this compound and interact very weakly with each other. Therefore,
the weak frequency dependence of the temperature corresponding to
the peak of $\chi_{\rm m}^\prime(T)$ in the samples with $x$ =
0.10 and 0.30 suggests existence of a cluster-glass, while the
frequency independence of the peak position for the sample with
$x$ = 0.60 indicates a ferromagnetic transition.

Figure 5 shows the ZFC 1/$\chi_{\rm m}(T)$ as the function of $T$
for the sample with $x$=0.45. The $\chi_{\rm m}(T)$ can be well
fitted with the Curie-Weiss law in the temperature range above
$T_{\rm c}$. Based on the fitting result, the effective magnetic
moment per cobalt ions $\mu_{\rm eff-Co}$ can be obtained by
subtracting the Gd$^{3+}$ contribution ($\mu_{\rm
eff}$(Gd$^{3+}$)= 7.94$\mu_{\rm B}$) from the total $\mu_{\rm
eff}$= $\sqrt{8C}$ ($C$, the Curie constant).\cite{Dekker} The
obtained $\mu_{\rm eff-Co}$ is 2.96$\mu_{\rm B}$, which is much
less than 3.67$\mu_{\rm B}$  in La$_{1-x}$Sr$_{x}$CoO$_3$ with the
same doping level,\cite{Goodenough} indicative of lower spin state
for Gd(Sr) compounds relative to La(Sr) compounds. The inset of
Fig. 5 shows the $M(H)$ loop for the sample with $x$=0.45 at 5 K,
exhibiting ferromagnetic behavior with spontaneous magnetization
and clear hysteresis with a paramagnetic component. The
non-saturating component comes from the corporate effect of large
paramagnetic signal from Gd$^{3+}$ ions and the cluster nature of
ferromagnetic state,\cite{Lashkareva,Kunhs,Hoch,Ghoshray} where
some fraction of the Co spins exists in the paramagnetic matrix.
It should be pointed out that the coercive field for this sample
at 5 K is as large as 2850 Oe, which is much larger than that in
La(Sr) compounds \cite{Wu,Goodenough} but comparable to Nd(Sr)
compounds.\cite{Stauffer} This pronounced coercive behavior is
believed to reflect the magnetic inhomogeneity and the formation
of ferromagnetic clusters.\cite{Lashkareva,Kunhs,Hoch,Ghoshray} We
note that in the inset of the Fig. 5 the samples with $x$ = 0.10
and 0.30 also show a small spontaneous magnetization, which can be
comparable with that observed in low doping La(Sr)
compounds.\cite{Wu,Goodenough} Such a small spontaneous
magnetization at 5 K and the weak frequency dependence of the
temperature corresponding to the peak in $\chi_{\rm m}^\prime(T)$
for the two samples indicate a cluster-glass. Actually,
Rey-Cabezudo et al.\cite{Cabezudo} have considered a cluster
picture about the magnetism for the Gd(Sr) samples with $x \leq$
0.30.

\subsection{Transport properties}

Figure 6 shows the temperature dependence of resistivity $\rho(T)$
for the Gd$_{1-x}$Sr$_x$CoO$_{3-\delta}$ system ($x$= 0.10 $\leq x
\leq$ 0.70). The $x$= 0.10 sample exhibits an insulating behavior
in the whole temperature range. With increasing Sr up to 0.45,
$\rho(T)$ decreases dramatically. At 4.2 K, $\rho(T)$ drops by
more than 1000 times for the sample with $x$= 0.30 and by more
than 10$^5$ times for the sample with $x$= 0.45 relative to the
sample with $x$=0.10. An insulator-metal transition occurs around
110 K as the Sr content increases up to 0.35. The $x$= 0.45 sample
shows metallic behavior down to 4.2 K and a kink around 120 K,
which coincides with the ferromagnetic transition temperature.
Such a kink in $\rho(T)$ is a common feature for an itinerant
ferromagnet because of the reduction of scattering from spin
disorder in ferromagnetic state as observed in CMR manganites
\cite{Tokura} and La$_{1-x}$Sr$_{x}$CoO$_3$ (0.30 $\leq~x
\leq~$0.60).\cite{Goodenough} With further increasing Sr content
above 0.45, the $\rho(T)$ increases rapidly. The $\rho(T)$ for the
sample with $x$= 0.50 shows still metallic behavior in whole
temperature range, while an upturn in low temperature is observed
in $\rho(T)$ for the sample with $x$= 0.55. It should be pointed
out that the position of the kink in $\rho(T)$ shifts to higher
temperature with increasing $x$ from 0.40 to 0.55, which is
consistent with the enhancement of $T_{c}$ with increasing $x$ as
shown in Fig. 3. As Sr content increases to 0.60, the resistivity
shows a reentrance of the insulating state.

Figure 7 shows the isothermal magnetoresistance at 20 K as a
function of magnetic field for Gd$_{1-x}$Sr$_x$CoO$_{3-\delta}$
with $x$=0.10, 0.45, and 0.60, respectively. A large negative MR
[($\rho(0)-\rho(H))/\rho(0)]\times100\%$ as high as -28.5$\%$ is
achieved in the $x$ = 0.10 sample at 13.5 T. The x$ = 0.45$
sample, which is the most metallic among all the samples, exhibits
a smallest negative MR $\approx$ -6$\%$ at 13.5 T. The MR in $x$=
0.60 sample increases to -14$\%$ at 13.5 T. This suggests that
magnetic field has the strongest effect on the most insulating
samples. It suggests that the MR depends on not only the
ferromagnetic state, but also the insulating state.

Attention should be paid to the insulator-reentering behavior for
$x \geq$ 0.60 since the end compound SrCoO$_3$ is metallic. To
consider the origin for the reentrance of the insulating state,
one must note the fact that it is difficult to achieve full oxygen
stoichiometry at high-doping level in Ln$_{1-x}$Sr$_x$CoO$_3$
system. Yo et al.\cite{Yo} found that there exists large oxygen
deficiency in Dy$_{1-x}$Sr$_{x}$CoO$_{3-\delta}$ and
Sm$_{1-x}$Sr$_{x}$CoO$_{3-\delta}$, and the oxygen deficiency
reaches 0.29 and 0.36 for the samples with $x$ = 0.75 for the two
systems, respectively. Ryu et al.\cite{Ryu} reported that there
also exists large oxygen deficiency in
Gd$_{1-x}$Sr$_{x}$CoO$_{3-\delta}$ for the samples with high
doping level. Therefore, in order to understand the reentrance of
the insulator state, the oxygen content is determined. The
K$_2$Cr$_2$O$_7$ titration experiments indicate that the oxygen
contents are 2.912, 2.728, and 2.602 for as-grown samples with $x$
= 0.45, 0.60, and 0.70, respectively, while 2.935, 2.745, 2.703
after annealing under the oxygen pressure of 165 atm at 500
$\celsius$ for 48h. It turns out that there really exists a large
oxygen deficiency at high doping level and annealing procedure
under high oxygen pressure reduces the deficiency. Compared to the
annealing procedure at 500 $\celsius$, the samples are annealed
under the high oxygen pressure of 240 atm at 900 $\celsius$, and
the resistivity for these samples is shown in fig. 8. The sample
with $x$ = 0.60 exhibits metallic resistivity in the whole
temperature range below 300 K. The sample with $x$ = 0.70 also
becomes much less insulating, with $\rho(T=5$ K$)$/$\rho(T=300$
K$)<$4/3, in contrast to that annealed at 500 $\celsius$ (larger
than 100). Metallic state should be restored in the sample with
$x$ = 0.70 after further annealing under higher oxygen pressure.
The K$_2$Cr$_2$O$_7$ titration experiments indicate that the
oxygen contents are 2.787 and 2.755 for $x$ = 0.60 and 0.70 after
annealing under the oxygen pressure of 240 atm at 900 $\celsius$,
respectively. It is clear that the oxygen deficiency leads to the
reentrance of the insulating behavior for the samples with $x$
$\geq$ 0.60.

\section{Discussion}
Figure 9 shows the phase diagram of
Gd$_{1-x}$Sr$_x$CoO$_{3-\delta}$ ($0.10 \leq x \leq 0.70$)
according to the above results. In Fig. 9, one can found a
pronounced feature that the Gd$_{1-x}$Sr$_x$CoO$_{3-\delta}$
system has much lower ferromagnetic transition temperature
compared to La(Sr) and Nd(Sr) compounds, for instance, $T_{\rm c}$
in Gd$_{0.5}$Sr$_{0.5}$CoO$_{3-\delta}$ annealed under the high
oxygen pressure of 240 atm at 900 $\celsius$ ($\delta \approx$
0.032) is around 150 K, while in La$_{0.5}$Sr$_{0.5}$CoO$_{3}$ it
is about 240 K. Another feature in Fig. 9 is that the critical Sr
concentration for MIT ($\approx$ 0.35) is much larger than that in
La(Sr) and Nd(Sr) compounds. Taking into account the fact that the
less conducting Gd$_{1-x}$Sr$_x$CoO$_{3-\delta}$ samples have the
lower $T_{\rm c}$, one may suppose a correlation between metallic
conductivity and ferromagnetic order as it would be expected with
a double-exchange model as in doped manganites. This opinion could
be supported by the fact that both the conductance and $T_{\rm c}$
increases as oxygen content increases, as inferred from Fig.6,
Fig. 8 and Fig. 9. It is well known that $T_{\rm c}$ in doped
manganites is argued to be mainly determined by two kinds of
structural distortion.\cite{Rodriguez} This is also believed to be
plausible in cobaltites.\cite{Kriener,Vanitha} The first is a
global distortion arising from the deviation of the structure from
the cubic symmetry, which is described by the deviation of
tolerant factor  $t$ (= $(<r_A>+r_{\rm O})/\sqrt{2}(r_{\rm
Co}+r_{\rm O})$ for the formation of ACoO$_3$). The much smaller
ion radius of Gd$^{3+}$ ion relative to La$^{3+}$ ($^{\rm
\@Roman{12}}r_{\rm La^{3+}}$= 1.36 ${\rm \AA}$) or Nd$^{3+}$
($^{\rm \@Roman{12}}r_{\rm Nd^{3+}}$= 1.30 ${\rm \AA}$)
ions\cite{Shannon,Shannon1} leads to a much smaller tolerant
factor in Gd(Sr) compounds compared to those in La(Sr) and Nd(Sr)
compounds. The second is a local distortion arising from the
different ion radii at A site, which is described by the variance
of the A-site ionic radii $\sigma^2$ (= $(1-x) r_{Ln}^2+x r_{M}^2
- <r_A>$ for Ln$_{1-x}$M$_{x}$CoO$_3$, where $<r_A>$ = $(1-x)
r_{Ln}+x r_{M}$.). The radius of  Sr$^{2+}$ ion is much larger
than that of Gd$^{3+}$ ion, which results in a large local
distortion (such as, when Ln$_{0.5}$Sr$_{0.5}$CoO$_3$, $\sigma^2$
= 0.0123 for Ln = Gd, while 0.0016 for Ln = La, and 0.0072 for Ln
= Nd).\cite{Vanitha} Consequently, the small tolerant factor in
Gd$_{1-x}$Sr$_x$CoO$_{3-\delta}$ samples means a large deviation
from cubic symmetry, and the large $\sigma^2$ suggests a
pronounced local disorder. They lead to a reduction in the
ferromagnetic exchange dramatically and thus the ferromagnetic
transition temperature and conductance. In addition,
Gd$_{1-x}$Sr$_x$CoO$_{3-\delta}$ is a more complex system due to
the large oxygen deficiency at high doping level, which would
influence the structural distortion and the carrier concentration
markedly. This could be another cause for the low ferromagnetic
transition temperature and conductance.

Actually, there is a further reason for the suppression of
conductance and ferromagnetism in
Gd$_{1-x}$Sr$_x$CoO$_{3-\delta}$. Compared to manganites, the
conversion of various spin-states of Co ions in cobaltites
influences the magnetic and the transport properties of cobaltites
seriously. The ferromagnetic exchange and charge transport are
thought to occur mainly through the hopping of $e_{\rm g}$
electrons, as shown in the inset of Fig. 9. The $t_{\rm 2g}$
electrons hopping can also occur, but it possesses a much smaller
possibility to take place. Therefore, the existence of $e_{\rm g}$
electrons is vital for metallic ferromagnetic order. Recently,
Lengsdorf et al.\cite{Lengsdorf} reported a transition from the
conducting state to the insulating state and a decrease of $T_{\rm
c}$ induced by pressure in La$_{0.82}$Sr$_{0.18}$CoO$_3$. This
peculiar behavior has been attributed to a gradual change of the
spin state for the trivalent ions from magnetic to a nonmagnetic
spin state under pressure. In LaCoO$_3$, it was found to undergo
an intermediate- to low-spin state transition under
pressure.\cite{Vogt} The change of the spin state with pressure is
realized due to the increase in the energy of the crystal-field
splitting ($\Delta_{\rm CF}$) under pressure. The increase of
$\Delta_{\rm CF}$ makes the low-spin Co$^{\rm \@Roman{3}}$ more
stable. Gd$^{3+}$ ion has a much smaller radius than La$^{3+}$,
therefore, the replacement of Gd$^{3+}$ for La$^{3+}$ ion has the
similar effect like a pressure applied to some extent. It
naturally leads to an increase of $\Delta_{\rm CF}$ and an
enhancement of the stability of the low-spin state than in the
La(Sr) compounds. Indeed, very recently Knizek et al. observed a
much larger $\Delta_{\rm CF}$ of Co ions in GdCoO$_3$ than that in
LaCoO$_3$.\cite{Kzinek} The much smaller effective magnetic moment
at Co ions in Gd(Sr) compounds relative to La(Sr) system,
suggested in Fig. 5, confirms the lower spin-state in Gd(Sr)
system. Furthermore, due to the higher acidity (i.e., a higher
charge/radius ratio) of Gd$^{3+}$, the Gd$^{3+}$ ions compete more
strongly with cobalt ions in covalent bonding to the oxygen atoms
than La$^{3+}$ ions. This leads to narrower Co-O bands and more
stable $\pi^{\ast}$(Co-O) levels.\cite{Goodenough2} This also
causes a more stable low-spin configuration in Gd(Sr) compounds.
Therefore, the smaller ion radius and the higher acidity of
Gd$^{3+}$ ion relative to La$^{3+}$ ion lead to a larger
$\Delta_{\rm CF}$, which favors a low-spin Co$^{\rm \@Roman{3}}$.
The stabilized low-spin Co ions results in the reduction of the
population of $e_{g}$ electrons. This is another important cause
for the less conductance and lower $T_{\rm c}$ in
Gd$_{1-x}$Sr$_x$CoO$_{3-\delta}$ relative to La(Sr) and Nd(Sr)
compounds.

Finally, the large magnetic moment of Gd$^{3+}$ (S = 7/2,
$\mu_{\rm eff}$ = 7.94 $\mu_{\rm B}$) should be considered. In
this paper, no obvious effect of the magnetic moment of Gd$^{3+}$
on magnetic and transport behavior is observed except for a strong
paramagnetic signal in low temperatures. Nevertheless, a larger
effective field than the applied field on Co ions system can be
achieved because the easy orientation of Gd$^{3+}$ sublattice in a
magnetic field. Rey-Cabezudo et al.\cite{Cabezudo} pointed out
that the paramagnetic Gd$^{3+}$ sublattice polarizes the cobalt
magnetic clusters. It has been considered as one possible reason
for the low-temperature MR. However, it has been reported that
large negative MR (more than 80$\%$ at 5 K for $x$=0.09) was
observed in low temperatures for insulator-like compositions of
La$_{1-x}$Sr$_{x}$CoO$_{3}$.\cite{Wu} Considering the fact of the
nonmagnetic La$^{3+}$ ion, the interpretation proposed by
Rey-Cabezudo et al. based on Gd$^{3+}$ ions may be in doubt. In
Ref. 24, Wu et al.\cite{Wu} interpreted such a negative MR at low
temperature in terms of a short-range FM ordered cluster model.
The Co$^{\rm \@roman{3}}$ and Co$^{\rm \@Roman{4}}$ in the
hole-rich clusters are aligned by magnetic field, so that an
increase in the electrons hopping possibility results in a
negative MR. Therefore, the smallest MR observed in most metallic
composition $x$ = 0.45 can be understood with the picture proposed
by Wu et al. in La(Sr) system. It should be pointed out that such
low-temperature smallest MR in the most conductive samples is also
observed in La(Sr) system.\cite{Wu} This indicates that the large
negative in low temperature in Gd(Sr) compounds has the same
origin as that in La(Sr) compounds.

\section{CONCLUSION}

The evolution of magnetic and transport properties with $x$ in
Gd$_{1-x}$Sr$_x$CoO$_{3-\delta}$ (0.10 $\leq x \leq$ 0.70)
annealed under the oxygen pressure of 165 atm at 500 $\celsius$
has been systematically investigated. Cluster-glass behavior is
observed in the low doping range, while the samples show the FM
transition in high doping region. The samples show insulator-like
behavior for $x \leq$ 0.30, and an IM transition occurs around $x$
= 0.35. The optimal doping is $x$ = 0.45. A striking feature is a
reentrance of insulator-like behavior in the samples with $x \geq$
0.60, arising form the large oxygen deficiency in the samples.
Annealing procedure under higher oxygen pressure at higher
temperature can restore the metallic state. Relative to the La(Sr)
and Nd(Sr) compounds, the Gd$_{1-x}$Sr$_x$CoO$_{3-\delta}$ system
exhibits less conduction and lower $T_{\rm c}$, which can be
attributed to the large global and local structural distortion,
and the more stable low-spin state of Co ions arising from the
small radius of Gd$^{3+}$ ion.

\section{ACKNOWLEDGEMENT}
\vspace*{-2mm}

This work is supported by the grants from the Nature Science
Foundation of China and by the Ministry of Science
and Technology of China, and the Knowledge Innovation Project of Chinese Academy of Sciences.\\

\clearpage

\begin{figure}[htbp]
\centering
\includegraphics[width=0.7\textwidth]{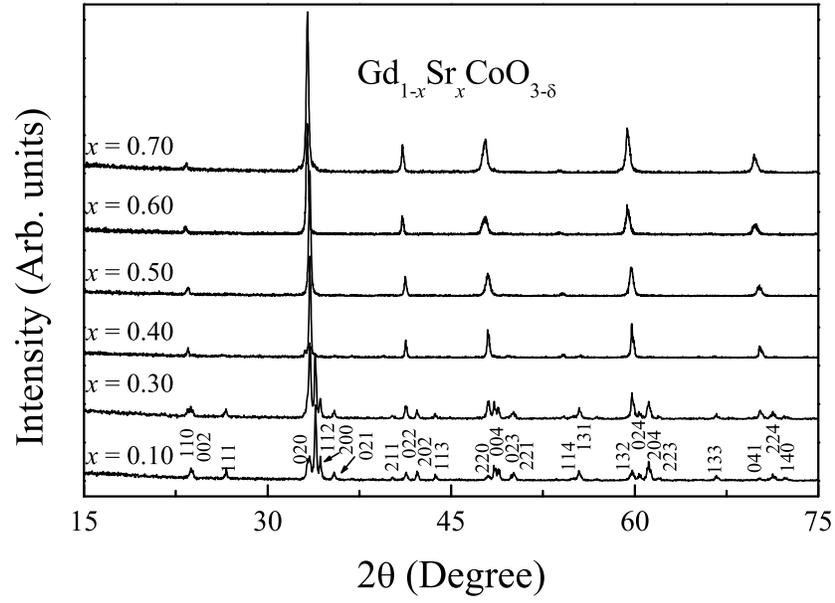}
\caption{The X-ray powder diffraction patterns for
Gd$_{1-x}$Sr$_x$CoO$_{3-\delta}$ ($x$= 0.10 $\leq x \leq$ 0.70). }
\label{Fig1}\vspace*{-2mm}
\end{figure}

\clearpage

\begin{figure}[htbp]
\centering
\includegraphics[width=0.7\textwidth]{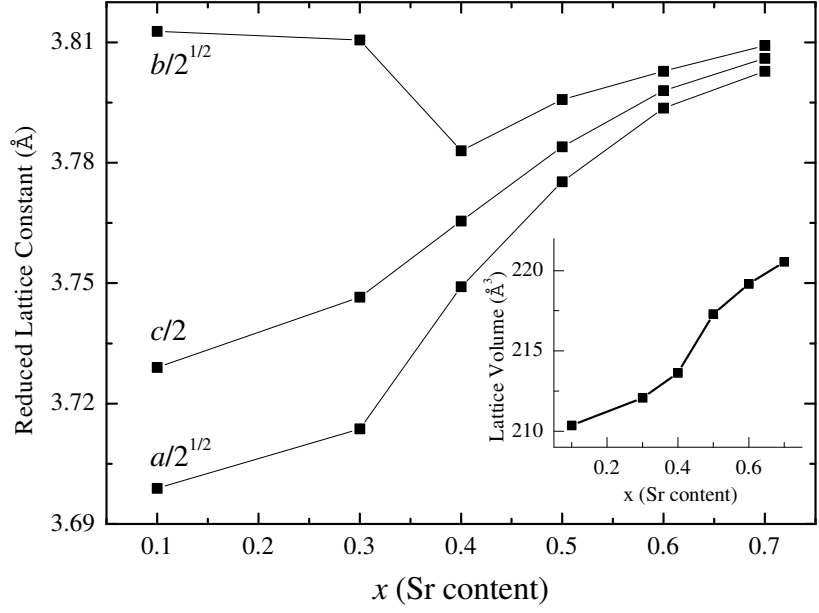}
\caption{The reduced lattice constants vs. Sr content ($x$) in
Gd$_{1-x}$Sr$_x$CoO$_{3-\delta}$ ($x$= 0.10 $\leq x \leq$ 0.70).
Inset: The variation of the lattice volume with Sr content ($x$).}
\label{Fig2}\vspace*{-2mm}
\end{figure}

\clearpage

\begin{figure}[t]
\centering
\includegraphics[width=0.7\textwidth]{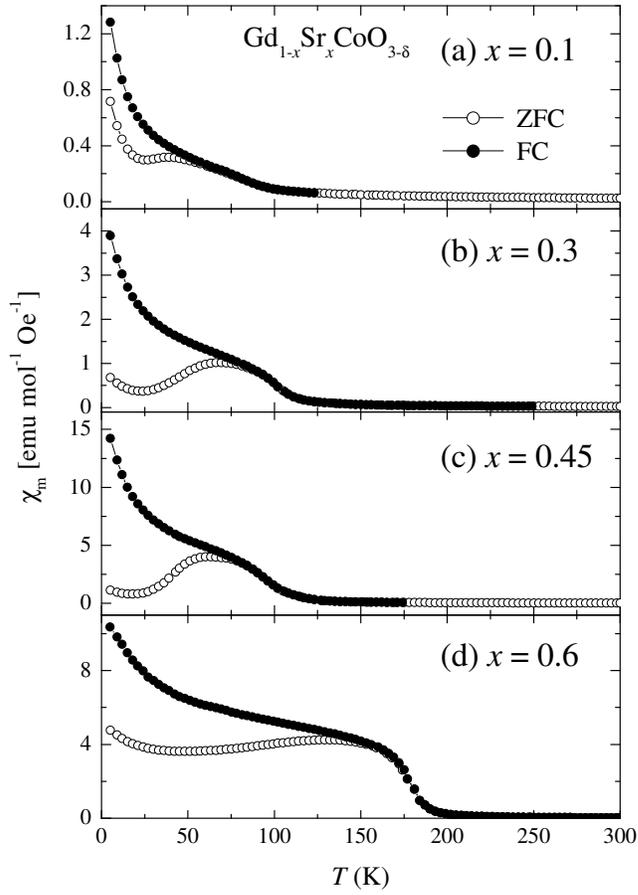}
\caption{Zero-field cooled (ZFC) and field-cooled (FC) molar
magnetic susceptibility of Gd$_{1-x}$Sr$_x$CoO$_{3-\delta}$ ($x$=
0.10, 0.30, 0.45, and 0.60) as a function of temperature at $H$=
1000 Oe.} \label{fig3}
\end{figure}

\clearpage

\begin{figure}[htbp]
\centering
\includegraphics[width=0.7\textwidth]{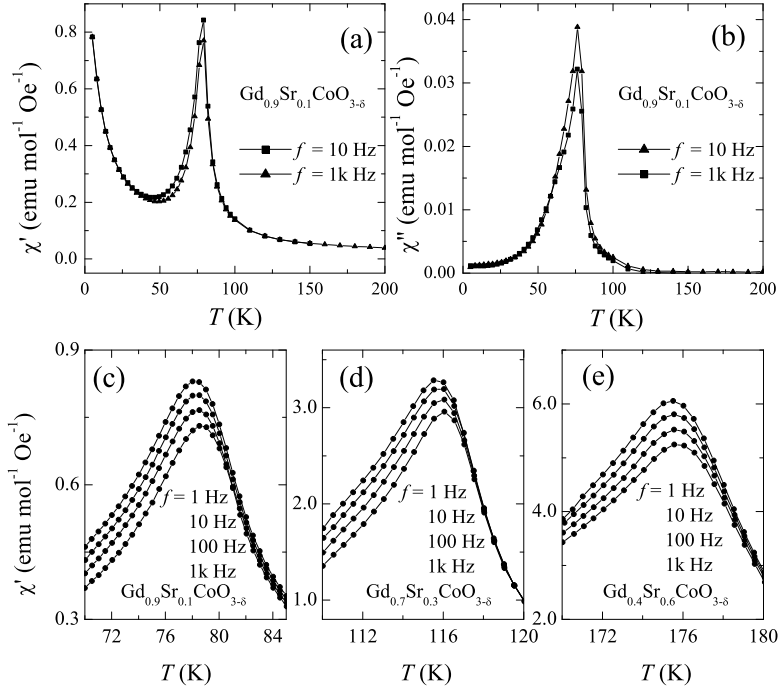}
\caption{(a),(b): Temperature dependence of the in-phase and
out-of-phase component of the ac susceptibility for $x$= 0.10. The
data were taken at 10 and 1000 Hz as indicated in the figure.
(c),(d),(e): Closeup of the temperature dependence of the in-phase
ac susceptibility  at the four frequencies in the range 1-1000 Hz
for the samples with $x$= 0.10 , 0.30  and 0.60, respectively.}
\label{fig4}
\end{figure}

\clearpage

\begin{figure}[htbp]
\centering
\includegraphics[width=0.7\textwidth]{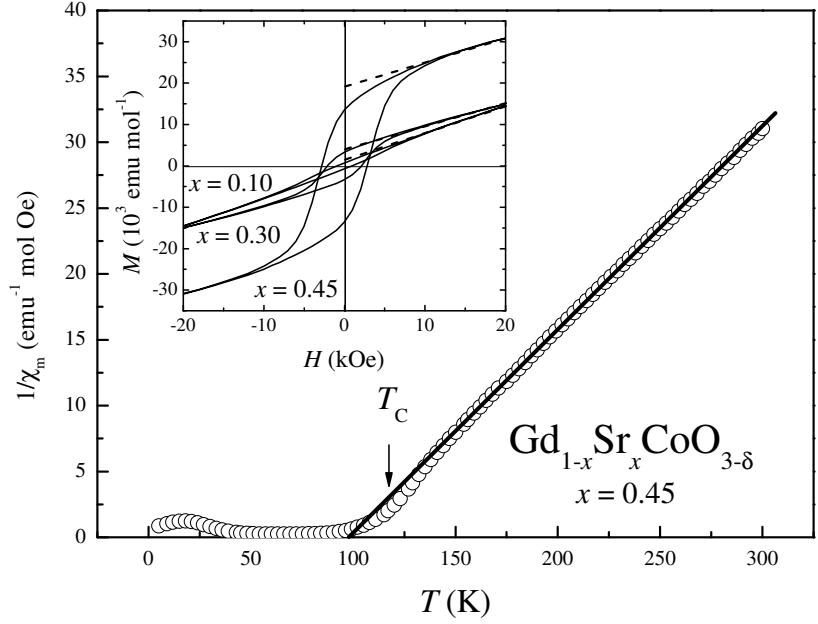}
\caption{The inverse ZFC molar magnetic susceptibility as a
function of temperature for the sample with $x$ = 0.45. The solid
line is the fitting curve according to Curie-Weiss law for the
high temperature data. The inset shows the $M$-$H$ loop for the
samples with $x$ = 0.10, 0.30 and 0.45 at 5 K between -2 T to 2 T.
The dash lines in the inset extrapolate the $M(H)$ to $H$=0 to
determined the spontaneous magnetization.} \label{fig5}
\end{figure}

\clearpage

\begin{figure}[htbp]
\centering
\includegraphics[width=0.7\textwidth]{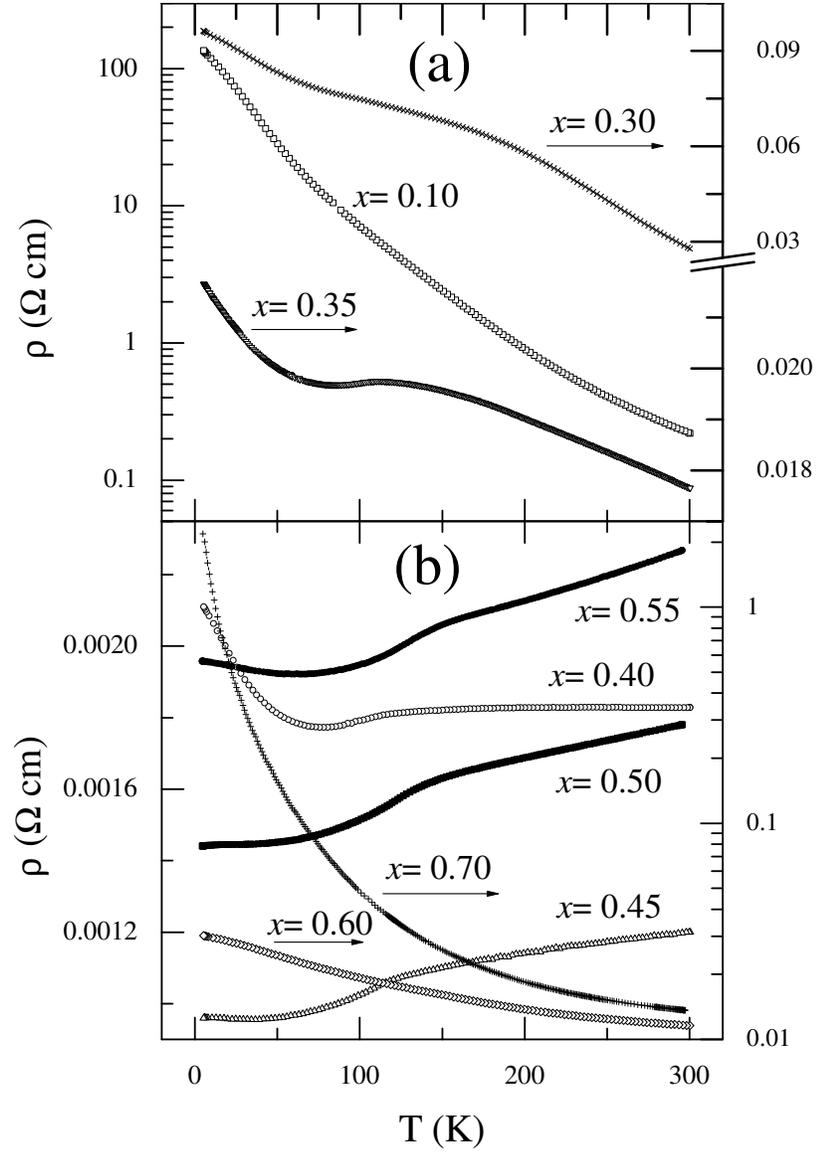}
\caption{The temperature dependence of resistivity for
Gd$_{1-x}$Sr$_x$CoO$_{3-\delta}$ ($x$= 0.10 $\leq x \leq$ 0.70).}
\label{fig6}
\end{figure}

\clearpage

\begin{figure}[htbp]
\centering
\includegraphics[width=0.7\textwidth]{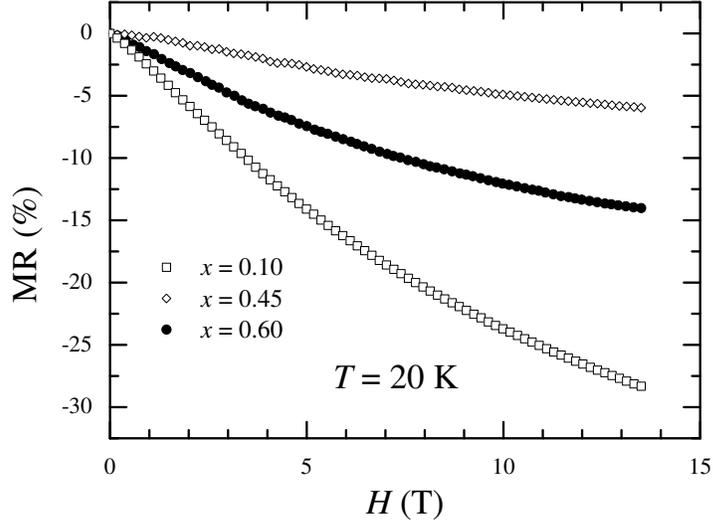}
\caption{The isothermal Magnetoresistance ($T$ = 20 K) as a
function of magnetic field for the samples with $x$=0.10, 0.45,
and 0.60, respectively.} \label{fig7}
\end{figure}

\clearpage

\begin{figure}[t]
\centering
\includegraphics[width=0.7\textwidth]{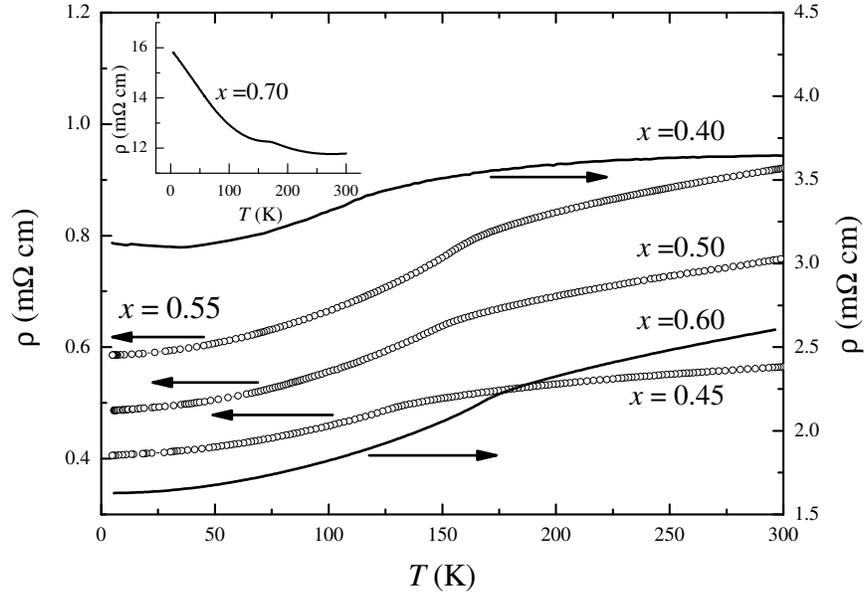}
\caption{The temperature dependence of the resistivity for the
samples with $x$=0.40-0.70 after annealing under the high oxygen
pressure of 240 atm at 900 $\celsius$ for 48 h.} \label{fig8}
\end{figure}

\clearpage

\begin{figure}
\includegraphics[width=0.7\textwidth]{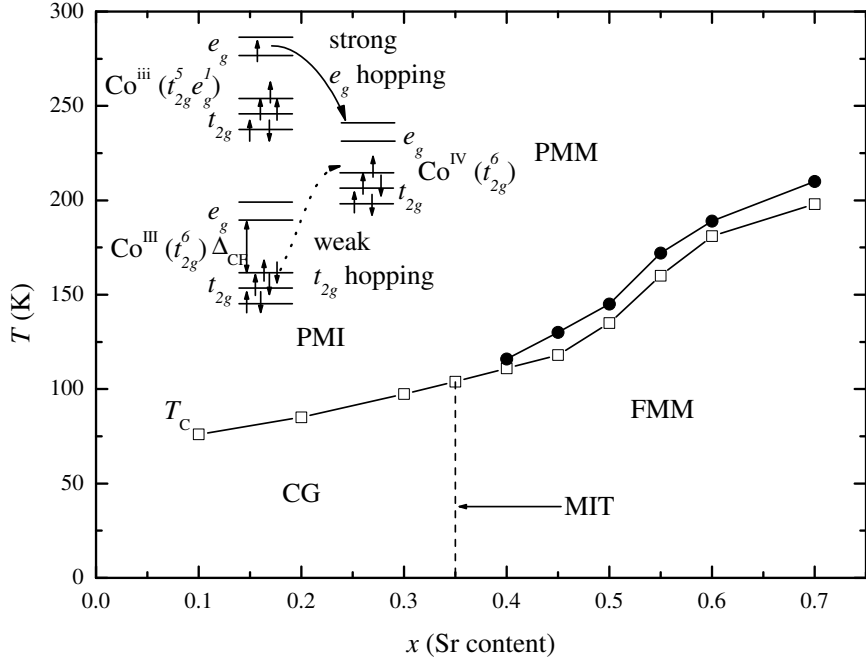}
\caption{The schematic drawing of the $T$-{\sl x} phase diagram
for the Gd$_{1-x}$Sr$_x$CoO$_{3-\delta}$ (0.10 $\leq x \leq$ 0.70)
system. $T_{c}$: determined from both dc susceptibility and ac
susceptibility ($\Box$: the samples annealed under the oxygen
pressure of 165 atm at 500 $\celsius$; $\bullet$: the samples
annealed under the oxygen pressure of 240 atm at 900 $\celsius$);
CG: cluster-glass; FMM: ferromagnetic metal; PMI/PMM: paramagnetic
insulator/metal; MIT: metal-insulator transition. Inset: possible
hopping procedure between trivalent cobalt ions and Co$^{\rm
\@Roman{4}}$. $\Delta_{\rm CF}$: the energy of crystal field
splitting.} \label{Fig9}
\end{figure}

\end{document}